\documentclass[twocolumn,showpacs,showkeys,preprintnumbers]{revtex4}

\def\be{\begin{equation}}
\def\ee{\end{equation}}
\def\beq{\begin{eqnarray}}
\def\eeq{\end{eqnarray}}
\def\n{\nonumber}
\def\bay{\begin{array}}
\def\eay{\end{array}}

\begin{document}

\preprint{CIRI/02-smw05}
\title{Spherical gravitational collapse and electromagnetic fields
in radially homothetic spacetimes}

\author{Sanjay M. Wagh}
\affiliation{Central India Research Institute, Post Box 606, Laxminagar, Nagpur 440 022, India\\
E-mail:cirinag@nagpur.dot.net.in}

\date{October 2, 2002}
\begin{abstract}
We consider a spherically symmetric, Petrov-type D, spacetime with
hyper-surface orthogonal, radial, homothetic Killing vector. In
this work, some general properties of this spacetime for
non-singular and non-degenerate data are presented. We also
present the source-free electromagnetic fields in this spacetime.
We then discuss general astrophysical relevance of the results
obtained for this spacetime.
\end{abstract}

\maketitle

\newpage

\section{Introduction}
In general relativity, non-gravitational processes are included
via the energy-momentum tensor for matter. Various
non-gravitational processes determine, apart from other physical
characteristics such as radiation, the relation of density and
pressure of matter. The temporal evolution of matter is to be
determined from such a relation, and from other physical
relations, if any.

By {\em physically realizable gravitational collapse}, we mean
gravitational collapse of matter that leads matter, step by step,
through different ``physical" stages of evolution, namely, from
dust to matter with pressure and radiation, to matter with
exothermic nuclear reactions etc.

Then, the spacetime of ``physically realizable" collapse of matter
must be able to begin with any stage in the chain of evolution of
matter under the action of its self-gravity. The temporal
evolution from any ``initial" data, any ``physical" stage in
question, is to be obtained from applicable non-gravitational
properties of matter. This is as per the principle of causal
development of data.

In a recent work \cite{cqg1}, we obtained a spherically symmetric
spacetime by considering a metric separable in co-moving
coordinates and by imposing a relation of pressure $p$ and density
$\rho$ of the barotropic form $p=\alpha\rho$ where $\alpha$ is a
constant. Such a relation determined only the temporal metric
functions of that spacetime.

Therefore, although the temporal metric functions for this
spacetime were determined in \cite{cqg1} using the above
barotropic equation of state, same metric functions are
determinable from {\em any\/} relation of pressure and density.
(See later.)

This spacetime admits a hyper-surface orthogonal, radial,
Homothetic Killing Vector (HKV). (See later.) Hence, it will be
called a {\em radially homothetic spacetime}. Since this spacetime
describes appropriate ``physical" stages of evolution of spherical
matter, we have argued in \cite{physical} that it is the spacetime
of {\em physically realizable\/} spherical collapse of matter. We
have also shown in \cite{physical} that it is a Petrov-type D
spacetime. We note that {\em all\/} general relativistic black
hole spacetimes are Petrov-type D spacetimes.

In \cite{sscollapse}, we studied the shear-free problem for the
sake of physical understanding of the issues involved in the study
of the spherical gravitational collapse in the spacetime of
\cite{cqg1}.

In this paper, we consider a radially homothetic spherically
symmetric problem for non-singular and non-degenerate matter data
in its full ``physical" generality, {\it ie}, with shear and
radiation. Further, we also obtain the source-free electromagnetic
fields in this spacetime explicitly.

We describe, in \S\ref{spacetime}, the spacetime under
consideration and its properties. Next, we summarize the
Hertz-Debye formalism in \S\ref{hdform}. The electromagnetic
fields in a radially homothetic spacetime are obtained in
\S\ref{eminrad}. In \S\ref{discussion}, we discuss the
astrophysical relevance and other implications of the results
obtained in these works.

\section{Spacetime metric} \label{spacetime}
In general, a HKV captures \cite{carrcoley} the notion of the
scale-invariance of the spacetime. If, in terms of the chosen
coordinates, a homothetic Killing vector ${\bf X}$ has component
only in the direction of one coordinate, the Einstein field
equations separate for that coordinate, generating also an
arbitrary function of that coordinate. This is the broadest (Lie)
sense of the scale-invariance leading not only to the reduction of
the field equations as partial differential equations to ordinary
differential equations but leading also to their separation.

A spherically symmetric spacetime has only one spatial scale
associated with it - the radial distance scale. Therefore, for a
{\em radially homothetic spacetime}, the metric admits one
arbitrary function of the radial coordinate. We then obtain for a
radially homothetic spacetime arbitrary radial characteristics for
matter. That is, due to the radial scale-invariance of the
spherical spacetime, matter has {\em arbitrary\/} radial
properties in a radially homothetic spacetime.

In co-moving coordinates, a radially homothetic, spherical
spacetime admits a spacelike HKV of the form \be
X^a\;=\;(0,\frac{y}{\gamma y'},0,0) \label{hkvradial} \ee and the
spacetime metric is given by \be
ds^2=-y^2dt^2+\gamma^2(y')^2B^2dr^2+y^2Y^2d\Omega^2
\label{ssmetfinal}\ee with a prime indicating a derivative with
respect to $r$, $B\equiv B(t)$, $Y\equiv Y(t)$ and $\gamma$ being
a constant. (We absorb the temporal function in $g_{tt}$ by
redefinition of the time coordinate.)

The Ricci scalar for (\ref{ssmetfinal}) is: \beq {\cal R} &=&
\frac{4\dot{Y}\dot{B}}{y^2YB}+\frac{2\ddot{B}}{y^2B}
-\frac{6}{y^2\gamma^2B^2}\n \\
&&\qquad\qquad  +\frac{2}{y^2Y^2} +\frac{2\dot{Y}^2}{y^2Y^2}
+\frac{4\ddot{Y}}{y^2Y} \label{ricciscalar} \eeq

The non-vanishing components of the Weyl tensor for
(\ref{ssmetfinal}) are: \beq \label{weyl} C_{trtr} &=&
\frac{B^2\gamma^2(y')^2}{3}\,F(t)  \\
C_{t\theta t\theta} &=& -\,\frac{y^2Y^2}{6}\,F(t)  \\ C_{t\phi
t\phi} &=& \sin^2{\theta}\,C_{t\theta t\theta}
\\ C_{r\theta r\theta} &=&\frac{B^2\gamma^2Y^2(y')^2}{6}\, F(t) \\ C_{r\phi
r\phi} &=& \sin^2{\theta}\,C_{r\theta r\theta}
\\ C_{\theta\phi\theta\phi} &=& -\,\frac{y^2Y^4\sin^2{\theta}}{3}\,F(t)  \eeq where
\be F(t) = \frac{\ddot{Y}}{Y} -\frac{\dot{Y}^2}{Y^2}-\frac{1}{Y^2}
-\frac{\ddot{B}}{B}+\frac{\dot{B}\dot{Y}}{BY} \ee

{\em In what follows, we shall assume, unless stated explicitly,
that there are no singular initial-data and that there are no
degenerate situations for the metric (\ref{ssmetfinal}).} See
later for the singularities and the degeneracies of the metric
(\ref{ssmetfinal}).

It is well-known that spherical spacetimes are either Petrov-type
O or Petrov-type D. Type-D spacetimes are not conformally flat. As
can be easily verified, the spacetime of (\ref{ssmetfinal}) is not
conformally flat for non-singular and non-degenerate data. That
the spacetime of (\ref{ssmetfinal}) is of Petrov-type D can also
be seen differently by verifying \cite{physical} that only $\Psi_2
= -\, C_{abcd}\ell^am^b\tilde{m}^cn^d$ is non-vanishing for the
metric (\ref{ssmetfinal}) where $\ell$, $n$, $m$ and $\tilde{m}$
are the Newman-Penrose  \cite{mtbh} tetrad vectors for it.

\subsection{Elementary flatness and center}
Now, the spacetime of (\ref{ssmetfinal}) is required, by
definition, to be locally flat at all of its points including the
center.

In the case of (\ref{ssmetfinal}), a small circle of coordinate
radius $\epsilon$ with center at the origin has circumference of
$2\pi\epsilon$. On the other hand, the circle has the proper
radius $\gamma\,y'\,\epsilon$. Then, requiring that the ratio of
the circumference to the proper radius of the circle to be $2\pi$
in the neighborhood of the origin, we obtain the condition for the
center to possess a locally flat neighborhood as \be
{y'|}_{r\,\sim\,0}\;\approx\;1/\gamma \label{conrzero} \ee This
condition must be imposed on any $y(r)$. With this condition,
(\ref{conrzero}), the HKV of metric (\ref{ssmetfinal}) is, at the
center, $y|_{r=0}\,\partial/\partial_r$.

Now,  $y(r)$ is the ``area radius" in (\ref{ssmetfinal}). When
$y|_{r=0} \neq 0$, the orbits of the rotation group $SO(3)$ do not
shrink to zero radius at the center for (\ref{ssmetfinal}).
Consequently, the center is not regular for (\ref{ssmetfinal})
when $y|_{r=0} \neq 0$ although the curvature invariants remain
finite at the center.

Also, when $y|_{r=0} = 0$, the center is regular for the spacetime
of (\ref{ssmetfinal}). But, the curvature invariants blow up at
the center, then.

It is well-known \cite{mcintosh} that the center and the initial
data for matter, both, are not {\em simultaneously\/} regular for
a spherical spacetime with hyper-surface orthogonal HKV.
Therefore, the spacetime of (\ref{ssmetfinal}) does not possess a
{\em regular\/} center and {\em regular\/} matter data,
simultaneously.

However, the lack of regularity of the center of
(\ref{ssmetfinal}) for non-singular matter data is understandable
\cite{physical} since the orbits of the rotation group do not
shrink to zero radius for every observer. It is a relative
conception and the co-moving observer of (\ref{ssmetfinal}) is not
expected to observe the orbits shrink to zero radius. (See also
later.)

\subsection{Singularities of spacetime}
Clearly, we may use the function $y(r)$ in (\ref{ssmetfinal}) as a
new radial coordinate - the area coordinate - as long as $y' \neq
0$. However, the situation of $y'=0$ represents a coordinate
singularity that is similar to, for example, the one on the
surface of a unit sphere where the analogue of $y$ is
$\sin{\theta}$ \cite{synge}. The curvature invariants do not blow
up at locations for which $y'=0$.

The genuine spacetime singularities of the strong curvature,
shell-focussing type exist when either $y(r)\,=\,0$ for some $r$
or when the temporal functions vanish for some $t=t_s$.

There are, therefore, two types of curvature singularities of the
spacetime of (\ref{ssmetfinal}), namely, the first type for
$B(t_s)=0$ and, the second type for $y(r)=0$ for some $r$.

Note that the ``physical'' radial distance corresponding to the
``coordinate'' radial distance $\delta r$ is \be \ell\;=\;\gamma
(y') B \delta r \ee Then, collapsing matter forms the spacetime
singularity in (\ref{ssmetfinal}) when $B(t)\,=\,0$ is reached for
it at some $t\,=\,t_s$. Therefore, the singularity of first type
is a singular hyper-surface for (\ref{ssmetfinal}).

The singularity of the second type is a singular sphere of
coordinate radius $r$. The singular sphere reduces to a singular
point for $r=0$ that is the center of symmetry. For $y(r)=0$ for
some range of $r$, there is a singular thick shell. Singularities
of the second type constitute a part of the initial data, singular
data, for the evolution.

\subsection{Degeneracies of the metric (\ref{ssmetfinal})}
The metric (\ref{ssmetfinal}) has evident degeneracies when
$y(r)=0$, $y(r)\to \infty$ either on a degenerate sphere of
coordinate radius $r$, for some ``thick shell" or globally. The
degeneracy $y(r)=0$ is equivalent to an infinite density while the
degeneracy $y(r)=\infty$ is equivalent to vacuum. (See later for
the expression of density of matter in the spacetime of
(\ref{ssmetfinal}).) Another degeneracy occurs for $y(r)=constant$
for some ``thick" shell or globally. This degeneracy corresponds
to uniform density.

\subsection{Self-similarity Nature}
Any vector (\ref{hkvradial}) can always be transformed
\cite{0112035} into \be \bar{X}^a\;=\;(T,S,0,0) \label{hkvusual}
\ee via a non-singular coordinate transformation \beq S &=& l(t)
\exp{\left(\int F^{-1} d r\right)} \n \\ \phantom{m}
\label{sstrans} \\ T &=& k(t) \exp{\left(\int F^{-1} d r\right)}
\n \eeq

We note that, if we invoke (\ref{sstrans}) for (\ref{ssmetfinal}),
the resulting metric will not be diagonal. The imposition of
diagonality of the metric will require a relationship between
$l(t)$ and $k(t)$. Such a relation can, of course, always be
imposed.

Hence, the metric (\ref{ssmetfinal}) can always be transformed,
under non-singular coordinate transformations (\ref{sstrans}), to
a form which admits a HKV (\ref{hkvusual}) in the transformed
coordinates. The transformed metric under consideration is,
therefore, \be ds^2\;=\;-\,P^2\,dT^2\; +\;Q^2\,dS^2\;+\;
S^2\,Z^2\, d\Omega^2 \label{ssmet} \ee where $P, Q, Z$ are the
metric functions of the self-similarity variable $T/S$ or $S/T$.
Note that for the transformed metric the radiation or the heat
flux is, in general, non-vanishing.

For (\ref{ssmet}), we are therefore led to consider the spacetime
singularity at $S=0$ and  $T=0$.

As we noted earlier, a relationship exists between $l(t)$ and
$k(t)$ of (\ref{sstrans}). Thus, $S=0$ {\it and} $T=0$ for
(\ref{ssmet}) corresponds to $l(t)=0$ for (\ref{ssmetfinal}) when
$l(t) \propto k(t)$ and $l(t)=0$. There is thus no constraint on
the radial function $y(r)$ in (\ref{ssmetfinal}) that it should
vanish. Consequently, $y(r) \neq 0$ at $r=0$ or, for that matter,
at any $r$, is permissible. Then, $y|_{r=0}$ is arbitrary.

However, $S=0$ and $T=0$ also corresponds to $y(r)=0$ in
(\ref{sstrans}) with $l(t)$ not proportional to $k(t)$. However,
assuming $y(r)=0$ leads to initially singular density. Such a
curvature singularity will then always exist on any hyper-surface
$t={\rm constant}$ in the spacetime of (\ref{ssmetfinal}).

We note that we may begin with the metric form (\ref{ssmet}) even
when the transformations (\ref{sstrans}) are singular. The metric
(\ref{ssmet}) is then not reducible to the form
(\ref{ssmetfinal}). For such a spacetime, the Einstein field
equations reduce to ordinary differential equations. However, the
resultant equations are not entirely separated in terms of the
variables $T$ and $S$.

\subsection{Mass function} \label{masstopic}
The mass function for the spacetime of (\ref{ssmetfinal}) can be
defined as \be m(r,t)
\;=\;\frac{yY}{2}\,\left(\,1\,-\,\frac{Y^2}{\gamma^2B^2}\right)
\label{mass} \ee where $m(r,t)$ denotes the effective total energy
per unit mass of a fluid element labelled by the co-moving radial
coordinate $r$ at co-moving time $t$. in the spacetime. Note that
it includes the ``effective'' contribution due to the flux of
radiation or heat in the spacetime.

Note that, using (\ref{mass}), we may rewrite the metric
(\ref{ssmetfinal}) as: \be
ds^2 = -\,d\tau^2 + \frac{R'^2\,dr^2}{1-(2\,m/R)
} + R^2d\Omega^2 \label{tbform} \ee where $R\equiv R(\tau, r) =
yY$, $m\equiv m(\tau, r)$ and $d\tau = y dt$. Now, the metric
(\ref{ssmetfinal}) is recognizable as a {\em generalization\/} of
the Tolman-Bondi dust metric to include pressure and radiation.

Now, we note the following. An asymptotic observer of the
Schwarzschild spacetime does not see a sphere, drawn around the
central mass point, shrink to a zero radius as a result of the
red-shift of the sphere becoming infinite at the gravitational
radius. Consequently, the ``area radius" of the sphere does not
shrink to a zero for the asymptotic observer but it appears to
this observer that the minimum area radius of the sphere is
$r=2M$, where $M$ is the central Schwarzschild mass. For the
asymptotic observer, $r=2M$ is then the ``center" of the
spacetime.

A co-moving observer, as a cosmological observer in
(\ref{ssmetfinal}), is ``equivalent" to the asymptotic observer of
the Schwarzschild geometry. Hence, for the co-moving observer of
(\ref{ssmetfinal}), the area of any sphere cannot shrink to zero
radius at the center since there is always an equivalent mass
point at the center for that observer. (See also \cite{physical}.)

Therefore, the lack of regularity of the center of
(\ref{ssmetfinal}) for non-singular matter data is understandable
\cite{physical} since the orbits of the rotation group do not
shrink to zero radius for the co-moving observer of
(\ref{ssmetfinal}). Note also that the singular hyper-surface of
(\ref{ssmetfinal}) is then the infinite-mass singularity.

\subsection{Field equations}
For completeness, we reproduce the arguments of \cite{physical}
here.

Now, define the co-moving time-derivative \be
D_t\;\equiv\;U^a\frac{\partial}{\partial x^a}
\;=\;\frac{1}{y}\frac{\partial}{\partial t} \ee where \be U^a =
\frac{1}{y}\;{\delta^{a}}_t \ee  is the four-velocity of the
co-moving observer. Then, the {\em radial\/} velocity of fluid
with respect to the co-moving observer is \be V^r
\;=\;D_t\left(yY\right)\;=\;\dot{Y} \ee  where an overhead dot has
been used to denote a time derivative.

The co-moving observer is accelerating for (\ref{ssmetfinal})
since \be {\dot U}_a = {U_a}_{;\,b}U^b\;=\;\left(\,0,\,
\frac{y'}{y},\, 0,\, 0\,\right) \label{acceleration}\ee is, in
general, non-vanishing for $y'\neq 0$. The expansion is \be \Theta
= \frac{1}{y}\,\left(\,\frac{\dot{B}}{B} \;+\;
2\,\frac{\dot{Y}}{Y}\,\right) \label{expansion} \ee

The Einstein tensor for (\ref{ssmetfinal}) is: \beq G_{tt}&=&
\frac{1}{Y^2}-\frac{1}{\gamma^2B^2} + \frac{\dot{Y}^2}{Y^2} +
2\frac{\dot{B}\dot{Y}}{BY}
\\ G_{rr}&=& \frac{\gamma^2B^2y'^2}{y^2} \left[-\,2\frac{\ddot{Y}}{Y}
-\frac{\dot{Y}^2}{Y} \right. \n \\ && \qquad \qquad \qquad \left.
+\frac{3}{\gamma^2B^2} - \frac{1}{Y^2}\right]
\\G_{\theta\theta}&=&-\,Y\,\ddot{Y}-Y^2\frac{\ddot{B}}{B}
- Y\,\frac{\dot{Y}\dot{B}}{B}+\frac{Y^2}{\gamma^2B^2}
\\G_{\phi\phi}&=& \sin^2{\theta}\,G_{\theta\theta} \\
G_{tr}&=&2\frac{\dot{B}y'}{By}  \label{gtr} \eeq

Now, to see, and only to see, that $\dot{B}$ is related to the
flux of radiation in the spacetime of (\ref{ssmetfinal}), we may
consider \cite{misner} matter to be described by the
energy-momentum tensor \be
\tilde{T}_{ab}\,=\,(\,p\,+\,\rho\,)\,U_a\,U_b \;+\; p \,g_{ab} \ee
and consider that it fails to satisfy a local conservation law due
to the emission of radiation that escapes radially along the
radial null vector $\ell^a$. For the radiation, we may then assume
the ``geometrical optics'' form \be E^{ab}\;=\;Q\,\ell^a\ell^b \ee
with $Q$ being the energy density of radiation or the energy flux
density in the rest frame of the fluid. Then, it is seen that $Q
\propto \dot{B}$. Thus, the radiation-flux depends on $\dot{B}$.

Now, define the quantity \be \sigma \equiv \,{{\sigma}^{1}}_{1} =
{{\sigma}^{2}}_{2} = - \frac{1}{2}{{\sigma}^{3}}_{3} =
\frac{1}{3y}\left( \frac{\dot{Y}}{Y}-\frac{\dot{B}}{B} \right)
\label{shearscalar} \ee Here, $\sigma_{ab}$ represents the
shear-tensor of the fluid and the shear-scalar is given by
$\sqrt{6}\;\sigma$. The spacetime of (\ref{ssmetfinal}) is
shearing when $B(t)$ is not proportional to $Y(t)$. Therefore, the
spacetime of (\ref{ssmetfinal}) is, in general, shearing and
radiating, both.

Now, we note that Penrose \cite{penroseweyl} is led to the Weyl
hypothesis on the basis of thermodynamical considerations, in
particular, those related to the thermodynamic arrow of time. On
the basis of these considerations, we may consider the Weyl tensor
to be ``some" sort of measure of the {\em entropy\/} in the
spacetime at any given epoch.

Then, for non-singular and non-degenerate data in
(\ref{ssmetfinal}), the Weyl tensor of (\ref{ssmetfinal}) blows up
at the singular hyper-surface of (\ref{ssmetfinal}) but is
``constant" at the ``initial" hyper-surface since $\dot{Y} =
\dot{B} =0$ for the ``initial" hyper-surface \cite{physical}.

This behavior of the Weyl tensor of (\ref{ssmetfinal}) is in
conformity with Penrose's Weyl curvature hypothesis
\cite{penroseweyl}. Thus, the spacetime of (\ref{ssmetfinal}) has
the ``right" kind of thermodynamic arrow of time in it.

\subsection*{Stages of collapse} Now, we turn to steps of
collapse of matter in the spacetime of (\ref{ssmetfinal}) for
non-singular and non-degenerate data. For non-singular and
non-degenerate data, we then have a ``cosmological" situation -
continued spherical collapse of matter from the assumed ``initial"
state.

\subsubsection*{Step I - Evolution of dust}
We begin with collision-less and pressureless dust matter with
density distribution given such that $y(r) > 1$ everywhere on the
initial hyper-surface in the spacetime of (\ref{ssmetfinal}).
Emission of radiation and, hence, radiation itself is not expected
in such dust.

Then, self-gravity leads to mass or energy-flux in the radial
direction. But, this is not the flux of radiation. Therefore,
there is no mass-flux in the rest frame of collapsing dusty
matter, but it is present for other observers in the spacetime.

Then, for vanishing flux of radiation, we have from (\ref{gtr}),
$B= {\rm constant}\equiv B_o$.

Then, the co-moving density, $\rho$, of dust is \be \rho
\,=\,\frac{1}{y^2}\,\left[ \frac{\dot{Y}^2}{Y^2} \, +
\frac{1}{Y^2}-\frac{1}{\gamma^2B_o^2} \right]
\label{dustdensity}\ee and the function $Y(t)$ is determined by
the condition of vanishing of the isotropic pressure: \be
4Y\ddot{Y}+\dot{Y}^2+1-\zeta\,Y^2 = 0 \ee Here, $\zeta =
5/\gamma^2B_o^2$, a positive constant.

A solution of this equation is obtainable as \be \frac{d
Y}{\sqrt{-1+\zeta/5 Y^2 + c_o Y^{-1/2}}} = t - t_0\ee where $c_o$
is constant. Since $\dot{Y}$ is the radial velocity of matter for
the co-moving observer, we require that solution for which
$\dot{Y}\to 0$ for $t\to -\infty$.

That the dust exists in the spacetime of (\ref{ssmetfinal}) is not
surprising since it is a generalization of the Tolman-Bondi dust
metric.

\subsubsection*{Step II - Evolution with pressure and radiation}
Next stage of collapse is reached when particles of dust begin to
collide with each other. Negligible amount of radiation, but
existing nonetheless, is expected from whatever atomic excitations
or from whatever free electrons get created in atomic collisions
in such matter. Therefore, dusty matter evolves into matter with
pressure and radiation, both {\em simultaneously} non-vanishing.

The energy-flux can no longer be removed by going to the rest
frame of matter.

Now, pressure and radiation, both, get {\em simultaneously\/}
switched on in the spacetime of (\ref{ssmetfinal}) when $B(t) \neq
0$. This is as per the expectation that dusty matter evolves to
one with simultaneous occurrence of pressure and radiation, both.

Then, the co-moving density, $\rho$ and isotropic pressure, $p$,
are given by \beq \rho &=&\frac{1}{y^2}\,\left[
\frac{\dot{Y}^2}{Y^2} \,+\,2\,\frac{\dot{Y}\dot{B}}{YB} +
\frac{1}{Y^2}-\frac{1}{\gamma^2B^2} \right]
\label{density} \\ \n \\
p &=& \frac{1}{y^2}\,\left[ \frac{5}{3\gamma^2B^2}-\frac{1}{3Y^2}
- \frac{4\ddot{Y}}{3Y} \right. \n \\
&&\qquad\qquad\quad \left.
-\,\frac{2\ddot{B}}{3B}-\frac{\dot{Y}^2}{3Y^2}-\frac{2\dot{Y}\dot{B}}{3YB}
\right] \label{pressure} \eeq

From (\ref{density}) and (\ref{pressure}), we obtain \be
2\,\frac{\ddot{Y}}{Y} +
\frac{\ddot{B}}{B}\;=\;\frac{2}{\gamma^2B^2}-\frac{y^2}{2}(\rho +
3p) \label{rddot} \ee

Then, from (\ref{rddot}), the relation of pressure and density of
matter is the required additional ``physical" information. Also
required is other relevant ``physical" information to determine
the radiation generation in the spacetime of (\ref{ssmetfinal}).
This is a non-trivial task in general relativity just as it is for
Newtonian gravity.

Note that the radiation may be ``negligible" but what is important
to considerations here is its presence in the spacetime.

\subsubsection*{Step III - Stellar object}
At some further stage of evolution, radiation from ``central part"
of collapsing matter may become non-negligible and the
temperatures in the central region may become appropriate for
exothermic, thermonuclear reactions. With the onset of exothermic
thermonuclear reactions, a ``shinning" star is born in the
spacetime.

The exothermic thermonuclear reactions in the stellar core may
support the overlying ``stellar layers" and such a stellar object
may ``appear" gravitationally stable.

But, the spacetime continues to be dynamic since radiation is
present in it. The central stellar object may also accrete matter
from its surrounding while emitting radiation.

Now, as and when ``heating" of the overlying stellar layers
decreases due to changes in exothermic thermonuclear processes in
the core of the star, the self-gravity of the stellar object leads
to its gravitational contraction. These are, in general, very slow
and involved processes.

Gravitational contraction leads to generation of pressure by
compression and by the occurrence of exothermic thermonuclear
reactions involving heavier nuclei. The star may stabilize once
more.

This chain, of gravitational contraction of star, followed by
pressure increase, followed by subsequent stellar stabilization,
continues as long as thermonuclear processes produce enough heat
to support the overlying stellar layers.

The theory of the atomic nucleus shows \cite{stars} that {\em
exothermic\/} nuclear processes do not occur when Iron nucleus
forms. With time, the rate of heat generation in
iron-dominated-core becomes insufficient to support the overlying
stellar layers which may then bounce off the iron-core resulting
into a stellar explosion, a supernova.

Then, many, different such, stages of evolution are the results of
physical processes that are unrelated to the phenomenon of
gravitation. These are, for example, collisions of particles of
matter, electromagnetic and other forces between atomic or
sub-atomic constituents of matter etc.

As an example, let some non-gravitational process, opposing
collapse, result into pressure that does not appreciably rise in
response to small contraction of the stellar matter. That is,
pressure does not appreciably rise when gravitational field is
increased by a small amount Then, the collapse of a sufficiently
massive object would not be halted by that particular
non-gravitational process. Therefore, a {\em mass limit\/} is
obtained in this situation. For example, electron degeneracy
pressure leads to the Chandrasekhar limit \cite{stars}.

Clearly, some of the non-gravitational processes determine the
gravitational stability of physical objects. This is true in
Newtonian gravity as well as in general relativity, both.

If we consider that density alone does not determine pressure
completely but that the isotropic pressure is a function of
density and temperature, both, then we need another equation, from
thermodynamical considerations, perhaps. In this case, the
relation of pressure and density can be considered to be a
function of time.

But, since the relation of pressure and density is {\em
arbitrary\/} for (\ref{ssmetfinal}), a changing pressure-density
relationship is allowed for it, we note. To provide for the
required physical information is, once again, a non-trivial task.

\vspace{.3in}

Evidently, to provide for the required information of ``physical"
nature is a non-trivial task in general relativity just as it is
for Newtonian gravity \cite{stars}. The details of these
considerations are, of course, very involved and have been left
out of the considerations of the present paper.

We note that, in general, the relation of pressure and density at
extremely high densities is not known. However, it can be surmised
that, in the final stages of collapse, the collapsing matter will
be ultra-relativistic and will end up in the singularity as such.
Then, the relation of pressure and density of such matter may be
expected to remain fairly unchanged in the final stages of the
gravitational collapse. That is to say, the spacetime of
collapsing matter should describe a relation of pressure and
density of ultra-relativistic matter that is not changing in the
final collapse stages.

Consequently, it seems reasonable to treat the thermodynamic state
of collapsing matter near the spacetime singularity by a relation
of the ``barotropic'' form $p=\alpha\rho$ where $\alpha$ is a
constant characteristic of the collapsing matter. Therefore, by
assuming this equation of state, we may obtain the temporal metric
functions in (\ref{ssmetfinal}) to study the gravitational
collapse in its final stages. But, for (\ref{ssmetfinal}), such a
possibility of final stage of collapse is approached only
asymptotically, that is for infinite co-moving time. (See \S
\ref{singhyper} later.)

But, it is clear that the field equations determine only the
temporal functions of (\ref{ssmetfinal}) from {\em any\/} suitable
energy-momentum tensor including that of electromagnetic fields,
if any.

The temporal functions $B(t)$ and $Y(t)$ are to be obtained from
the properties of matter such as a relation of pressure and
density, the rate of loss of internal energy to radiation,
processes of quantum mechanical nature etc.

Moreover, it is also clear that matter will continue to pile up on
such a star in a ``cosmological setting" and, hence, such a star
will always be taken over any mass-limit in operation at any stage
of its evolution. The evolution of collapsing matter will,
asymptotically, lead to the singular hyper-surface of the
spacetime of (\ref{ssmetfinal}).

The radial dependence of matter properties is ``specified" as
$1/y^2$ but the field equations of general relativity do not
determine the metric function $y(r)$ in (\ref{ssmetfinal}).

Therefore, the radial distribution of matter is {\em arbitrary} in
terms of the co-moving radial coordinate $r$. This is the
``maximal" {\em physical\/} freedom compatible with the assumption
of spherical symmetry, we may note. Note, however, that the
physical generality here is not be taken to mean the
``geometrical" generality.

\subsection{Absence of null or one-way membrane}
A spherical surface $r={\rm constant}$ in the geometry of
(\ref{ssmetfinal}) has a normal vector \be n_a=(0,1,0,0) \qquad
n^an_a\,=\frac{1}{\gamma^2 (y')^2B^2} \ee  Within the range $(0,
\infty)$ of the co-moving radial coordinate $r$, the character of
$n^a$ does not change from spacelike to null to timelike in the
spacetime of (\ref{ssmetfinal}). Then, the norm $n^an_a$ does not
vanish at any $r$ in non-singular and non-degenerate cases.

To see the same differently, the coordinate speed of light in the
spacetime of (\ref{ssmetfinal}) is \be \frac{dr}{dt}
\;=\;\pm\,\frac{y}{\gamma (y') B} \label{cspeed} \ee This speed
cannot vanish for non-singular and non-degenerate cases. At the
singular hyper-surface, the coordinate speed of light becomes
infinite.

Therefore, in the spacetime of (\ref{ssmetfinal}), there cannot be
a spherical, spatially finite, null membrane or a one-way
membrane, {\it ie}, a black hole in the usual sense of the term,
in non-singular and non-degenerate cases.

\subsection{Spherical collapse, singular hyper-surface and the
infinite red-shift surface} \label{ssaccrete} We emphasize that we
are using non-singular and non-degenerate data for
(\ref{ssmetfinal}). A co-moving observer in (\ref{ssmetfinal}) is
then ``a cosmological observer''. The four-velocity of matter
fluid with respect to a co-moving observer is: \be
u^a\;=\;\left(\,u^t,\,u^r,\,0,\,0\,\right) \qquad \qquad V^r =
\frac{u^r}{u^t} \ee We then obtain from the metric
(\ref{ssmetfinal}):
\beq u^a &=& \frac{1}{y\,\sqrt{\Delta}}\,\left(\,1,\,V^r,\,0,\,0\,\right) \\
\Delta &=&\;1\;-\;\gamma^2\,\left(
\frac{y'}{y}\right)^2\,B^2\,(V^r)^2 \label{Delta} \eeq

Now, if $d\tau_{\scriptscriptstyle CM}$ is a small time duration
for a co-moving observer and if $d\tau_{\scriptscriptstyle RF}$ is
the corresponding time duration for the observer in the rest frame
of matter, then we have \be d\tau_{\scriptscriptstyle
CM}\;=\;\frac{d\tau_{\scriptscriptstyle RF}}{\sqrt{\Delta}}
\label{rshift} \ee From (\ref{rshift}), we also get the red-shift
formula \be \nu_{\scriptscriptstyle CM} = \nu{\scriptscriptstyle
RF}\,\sqrt{\Delta}\qquad 1+z = 1+\frac{\nu{\scriptscriptstyle
RF}}{\nu{\scriptscriptstyle CM}}\ee in the spacetime of
(\ref{ssmetfinal}) where $\nu_{\scriptscriptstyle CM}$ is
frequency of a photon in the co-moving frame,
$\nu_{\scriptscriptstyle RF}$ is the frequency in the rest frame
and $z$ is the red-shift of the photon. Then, $\Delta\;=\;0$ is
the {\em infinite red-shift surface\/} that is, however, {\em
not\/} a null membrane. A co-moving observer waits for an infinite
period of its time to receive a signal from the rest-frame
observer when $\Delta\,=\,0$.

Then, we distinguish regions of the spacetime of
(\ref{ssmetfinal}) as \beq
(\Delta\,>\,0)\qquad |\gamma\,(y')\,B\,V^r| \;<\;y \label{bh0} \\
(\Delta\,=\,0)\qquad |\gamma\,(y')\,B\,V^r| \;=\;y \label{bh1} \\
(\Delta\,<\,0)\qquad |\gamma\,(y')\,B\,V^r| \;>\;y \label{bh2}
\eeq

Now, the geodesic equations of motion for (\ref{ssmetfinal}) are
easily obtainable. The $r$-equation is: \be \frac{d}{ds} \left(
\,\gamma^2B^2\,y\,y'\,\tilde{r} \,\right)\;=\;2{\cal L} \ee where
an overhead tilde denotes derivative with respect to the affine
parameter $s$. Then, \be \tilde{r}\;=\;\frac{2{\cal
L}\,s\,+\,k_1}{\gamma^2B^2\,y\,y'} \label{rgeoeq} \ee where $k_1$
is a constant of integration.

For the motion of the particle in the equatorial plane
$\theta=\pi/2$, the solution of the $t$--equation of motion, using
(\ref{rgeoeq}) in the lagrangian of motion, is \be
\tilde{t}=\sqrt{-\frac{2{\cal L}}{y^2}+\frac{1}{y^4}\left[
\frac{(2{\cal L}s+k_1)^2}{\gamma^2B^2}\right]} \ee

Then, since $V^r \equiv dr/ydt$, we have \be V^rV_r =
\frac{(k_1-s)^2}{y^2\gamma^2B^2 + (k_1-s)^2} \qquad(2{\cal
L}=-1)\ee where $s$ is the affine parameter along the geodesic and
we have $V^r = 0$ at $s=k_1$.

\subsubsection{Singular hyper-surface} \label{singhyper} Clearly, $V^rV_r=1$ for
$B(t_s)=0$ at $t=t_s$, {\it ie}, the velocity of the particle with
respect to a co-moving observer is the speed of light at the
singular hyper-surface of (\ref{ssmetfinal}) - $B(t_s)=0$ at
$t=t_s$. This is, of course, happening only asymptotically.

But, from (\ref{Delta}), $\Delta =1$ for $B(t_s)=0$. A co-moving
observer also moves with the speed of light at the singular
hyper-surface. After all, matter everywhere should become
relativistic as the central mass condensate continues to grow (due
to accretion onto it) to influence the entire spacetime to become
relativistic everywhere. This too is happening only
asymptotically. But, the singular hyper-surface is not the
infinite red-shift surface.

\subsubsection{Infinite red-shift surface}\label{bhole}
A spatially bounded infinite red-shift surface ``occurs" in
(\ref{ssmetfinal}) when $\Delta=0$. Since \be \Delta =
\frac{\gamma^2y^2B^2+(k_1-s)^2(1-\frac{1}{y^2})}{\gamma^2y^2B^2 +
(k_1-s)^2} \label{irseq} \ee this requires $y(r) < 1$.

Thus, the initial ``density distribution", from $y(r)$, decides
whether (\ref{ssmetfinal}) has an infinite red-shift surface or
not, {\em ie}, whether $y^2=V^rV_r$. This is possible for
$V^rV_r<<1$, but for sufficient mass concentration at the center,
{\em ie}, for $y < 1$.

But, a co-moving observer does not see the infinite red-shift
surface form. Thus, an initially ``small" matter density ($y > 1$)
cannot, for a co-moving observer, become ``large" enough that the
infinite red-shift surface forms ($y < 1$).

Matter piles up at the center of the spacetime. But, the density
at the center does not become infinite at any finite co-moving
time.

Now, let $y<1$, {\it ie}, let there be an infinite red-shift
surface in (\ref{ssmetfinal}) at the initial epoch itself. (For
$y|_{r=0}=0$, there is an infinite-density singularity at the
center.) Then, from (\ref{rgeoeq}), it follows that a particle can
reach and cross this infinite red-shift surface along its radial
geodesic.

But, such a central region - ``$y< 1$ apparition" - cannot
communicate to its exterior since the co-moving time duration is
imaginary in this region. Therefore, the interior of an infinite
red-shift surface is causally disconnected from the rest of the
spacetime at the initial time itself.

Then, matter external to $y< 1$ region eventually collapses onto
this central ``apparition" without a co-moving observer seeing
matter enter it. The spacetime of (\ref{ssmetfinal}) then
describes the accretion of matter onto the central ``apparition" -
the spatially bounded infinite red-shift surface.

{\it We shall refer to the infinite red-shift surface, as
described above, as the black hole surface and its interior as an
interior of a black hole}. This is the conception of a {\em black
hole\/} that arises in the spacetime of (\ref{ssmetfinal}) for
non-singular and non-degenerate data. But, here, a black hole is
{\em not\/} a null membrane or horizon.

\subsubsection{Apparent horizon} \label{light-trap} A radially
outgoing null vector of (\ref{ssmetfinal}) is \be
\ell^a\partial_a\;=\frac{1}{y}\frac{\partial}{\partial t}
\;+\;\frac{1}{\gamma y' B}\frac{\partial}{\partial r}\label{rnull}
\ee Light gets trapped inside a particular radial coordinate $r$
when the expansion of the above principle null vector vanishes at
$r$. The formation of the outermost {\em light-trapping surface\/}
or the {\em apparent horizon\/} is then obtained by setting the
expansion of (\ref{rnull}) to zero.

The zero-expansion of (\ref{rnull}) yields a condition only on the
temporal metric functions as \be
\frac{\dot{B}}{B}\,+\,2\,\frac{\dot{Y}}{Y}\; =\; -\frac{3}{\gamma
B} \label{ltscon} \ee The condition (\ref{ltscon}) implies an
``instant of time".

This is seen as as follows. An outgoing photon moves along the
trajectory
\[ \frac{dr}{dt} = \frac{y}{\gamma (y')B} \] and crosses a sphere
of coordinate radius $r$ at co-moving time $t$.

The mass inside this sphere is given by (\ref{mass}). Now, the
equation of the light trajectory can be used to express the mass
function $m$ as a function of either $r$ alone or $t$ alone. The
light trapping mass is then obtained by setting $2m = yY$. Then,
for every value of $r$ there is some $t$ and vice versa for which
$2m = yY$. Thus, condition (\ref{ltscon}) implies, in essence, an
``instant of light trapping''. Essentially, for non-singular and
non-degenerate data in (\ref{ssmetfinal}), every instant, of the
co-moving time, is an instant of light trapping.

Alternatively, let a spherical light front be emitted from the
center of symmetry. As it travels radially outwards, it brings in
more mass to its interior. When sufficient mass is in the
interior, light trapping occurs. For non-singular and
non-degenerate data in (\ref{ssmetfinal}), we can always draw a
sphere containing enough mass that can trap light. This is the
essence of the statement that ``Every co-moving instant is a
`Light Trapping Instant' in this spacetime". In a sense, every
observer is {\em inside\/} some light trapped sphere in
(\ref{ssmetfinal}) for non-singular and non-degenerate data.

\subsection{Shell black hole} Now, an interesting
possibility is that of (\ref{ssmetfinal}) containing many
concentric infinite red-shift surfaces. Then, what we have here is
the possibility of shells of black holes!

Intuitively, this is the only possibility that can arise in
spherical symmetry apart from that of a single spherical black
hole as considered earlier. It is also clear that this possibility
arises only as ``initial data" in the spacetime of
(\ref{ssmetfinal}).

To analyze such a ``shell" black hole, we will, of course, be
required to use the gaussian coordinate system since $y'=0$ at
some radial locations. We recall that $y'=0$ is a coordinate
singularity of the metric (\ref{ssmetfinal}).

That the shell black hole is obtained in the spacetime of
(\ref{ssmetfinal}) is not a coincidence since (\ref{ssmetfinal})
is the spherically symmetric spacetime with maximal ``physical"
freedom.

\subsection{When is the singularity of (\ref{ssmetfinal}) a locally or
globally naked singularity?} The radial null geodesic in
(\ref{ssmetfinal}) satisfies (\ref{cspeed}), {\it ie}, \[
\frac{dt}{dr}=\pm\, \gamma \frac{y'}{y} B \] The above
differential equation of the radial null geodesic does not possess
a singularity for non-singular and non-degenerate data.

For non-singular and non-degenerate data, the spacetime
singularity develops as a result of only the temporal evolution of
matter in the spacetime to the future, {\it ie}, as $B(t)\to 0$.

But, then the above tangent is vanishing at the singularity
indicating that there does not exist a future-directed, timelike
tangent at the spacetime singularity. Hence, {\em the curvature
singularity of (\ref{ssmetfinal}) is neither globally nor locally
visible for the non-singular and non-degenerate data.} It is a
singular hyper-surface occurring to the future of every observer
in the spacetime of (\ref{ssmetfinal}).

On the other hand, for singular data, {\it ie}, when $y(r)=0$ for
some $r$, the equation of radial null geodesic (\ref{cspeed}) has
a singularity at that $r$.

Then, whether such a singularity or singular sphere is visible to
any observer or not depends on the limit of the quantity $y'/y$ as
we approach the singular point for $B(t)\neq 0$. For some
functions $y(r)$ such a limit can be positive making the
singularity in question a locally naked or a locally visible one.

Therefore, for the spacetime of (\ref{ssmetfinal}), the visibility
of the spacetime singularity is determined principally by whether
we assume the existence of a visible singularity at the initial
time or not. The temporal evolution of non-singular and
non-degenerate data does not lead to a visible singularity in this
spacetime, we note.

A short comment on the the possible naked singularity solutions of
general relativity will not be out of place here.

For non-singular and non-degenerate data in (\ref{ssmetfinal}), no
naked singularities arise. A black hole, an infinite red-shift
surface, can exist in (\ref{ssmetfinal}) as a part of this initial
data.

Now, naked singularities can ``arise" for singular and degenerate
data in (\ref{ssmetfinal}). We emphasize that for singular and
degenerate data we have to abandon the metric (\ref{ssmetfinal})
and seek an entirely different solution of the field equations of
general relativity. In cases wherein the HKV of these spacetimes
of naked singularities is (\ref{hkvusual}) and it is not reducible
to (\ref{hkvradial}), the rationale of ``pure" radial
self-similarity is lost for these spacetimes.

In the absence of ``pure" radial scale-invariance, it is then not
clear what is the ``physical" significance of the existence of the
HKV (\ref{hkvusual}). Perhaps, there is none.

\section{Hertz-Debye formalism} \label{hdform}
Having presented the properties of the spacetime of
(\ref{ssmetfinal}), we now turn to the Hertz-Debye formalism
before embarking upon the nature of the electromagnetic fields in
the spacetime of (\ref{ssmetfinal}).

\subsection{Hertz-Debye potentials}
In flat space, Hertz \cite{hertz} introduced two vector potentials
$\vec{P}_E$ and $\vec{P}_M$ related to the standard
electromagnetic potentials $\Phi$ and $\vec{A}$ as \be \Phi = -\,
\vec{\nabla} \bullet \vec{P}_E \qquad \vec{A} =
\frac{\partial\vec{P}_E}{\partial t} + \vec{\nabla} \times
\vec{P}_M \ee and, hence, the bi-vector or the anti-symmetric
second-rank tensor potential is related by second derivatives to
the physical fields.

The gauge freedom associated with the Hertz potentials is such as
to preserve the source-free character of the Maxwell equations
while the gauge terms appear as sources in the equations: \be
\vec{Q}_E = \vec{\nabla}\times \vec{G} \qquad \vec{Q}_M =
-\,\frac{\partial\vec{G}}{\partial t} - \vec{\nabla}g \ee and \be
\vec{R}_E = -\,\frac{\partial\vec{W}}{\partial t} - \vec{\nabla}w
\qquad \vec{R}_M = -\,\vec{\nabla}\times \vec{W} \ee where
$(\vec{G}, g)$ and $(\vec{W}, w)$ are arbitrary 4-vectors.

This gauge freedom can be used \cite{nisbet} to reduce the Hertz
bi-vector to purely radial vectors of the form $\vec{P}_E =
P_E\,\hat{r}$ and $\vec{P}_M = P_M\,\hat{r}$ where $\hat{r}$ is
the unit radial vector. The functions $P_E$ and $P_M$ are the
Debye potentials \cite{debye} and obey, both, a wave equation. It
should be noted that only the ``monopole" field is missing in this
scheme \cite{cohen}.

In essence, the arbitrary, source-free electromagnetic field is
specified by two scalar functions which obey a single, separable
second-order wave equation. Therefore, a remarkable economy is
achieved by the Debye potentials. In \cite{cohen}, this is
expressed as: ``since a zero rest-mass field possesses two degrees
of freedom, no more economical representation of the Maxwell field
is possible" than that provided by the Debye potentials.

\subsection{Cohen \& Kegeles generalization}
Differential forms generalize the flat space Maxwell equations to
any curved spacetime in a natural way. Define the Maxwell 2-form
as \be f = \frac{1}{2}\;f_{ab}\, \omega^a \wedge \omega^b
\label{mform} \ee where $f_{ab}$ is the Maxwell tensor and
$\omega^a$ are the basis forms. The Maxwell equations are simply
\be df =0 \qquad\qquad \delta f =0 \label{meq} \ee where $d$ is
the exterior derivative and $\delta = \star \,d \star$ is the
co-derivative. Here, $\star$ is the Hodge dual operation.

The Hertz bi-vector $P$ (2-form) is related to the
electro-magnetic four-potential 1-form $A$ and the Maxwell 2-form
as \be A = \delta P \qquad \qquad f = d\delta P = - \,\delta d P
\label{afrelp} \ee Then, the equality of the last two expressions
in (\ref{afrelp}) requires that \be \triangle P \equiv (d\delta +
\delta d)P \equiv (d\star d\star + \star d\star d) P= 0 \ee where
$\triangle$ is the harmonic operator.

The 2-form gauge terms are: \be Q = dG \qquad \qquad R = \star
\,dW  \label{gauge} \ee where $G$ and $W$ are arbitrary 1-forms.
Therefore, the wave equation with the gauge terms is \be \triangle
P = dG + \star\, dW \label{waveeq} \ee so that the transformed
fields are \be f = d\delta P - dG = \star\, dW - \delta dP
\label{fields} \ee The transformed fields still obey the
source-free Maxwell equations as a consequence of the important
identities: \be d^2 \equiv \delta^2 \equiv 0 \ee resulting to
$df=0$ and $\delta f =0$.

Equations (\ref{waveeq}) and (\ref{fields}) provide an elegant and
fully covariant generalization of the Hertz potential formalism to
curved spacetimes.

The problem now consists of determining special bi-vector
directions in the spacetime so that (\ref{waveeq}) yields
decoupled wave equations for the corresponding components of the
potential for some choice of the gauge terms (\ref{gauge}).

In a class of spacetimes, the principal directions of the Weyl
tensor \cite{mtbh, npformalism} provide such special bi-vectors
\cite{cohen}. Such special bi-vector directions are defined
geometrically and independently of the Maxwell fields to be
computed. In essence, one chooses a null tetrad (the Carter
tetrad) with one null vector aligned along the repeated principal
null direction of the Weyl tensor of an algebraically special
spacetime in this scheme.

This completes our overview of the Hertz-Debye formalism  or the
Cohen-Kegeles formalism \cite{cohen}. We now turn to the problem
of electromagnetic fields in the spacetime of (\ref{ssmetfinal}).

\section{Electromagnetic fields in the spacetime of
(\ref{ssmetfinal})}\label{eminrad} Choosing an orthonormal tetrad
as a basis, the spacetime metric (\ref{ssmetfinal}) is \be ds^2 =
-\,(\omega^0)^2+(\omega^1)^2 +(\omega^2)^2 +(\omega^3)^2 \ee where
\beq \omega^0 =& y dt \qquad\qquad \omega^1 =& \gamma y' B dr \\
\omega^2 =& yY d\theta  \qquad\quad \omega^3 =& yY\sin{\theta}
d\phi \eeq The Hertzian potential 2-form is chosen to be \be P =
P_E\, \omega^0 \wedge \omega^1 + P_M\, \omega^2 \wedge \omega^3
\label{bvec}  \ee Under the Hertz-Debye formalism, we obtain a
wave equation for $P_E$ and $P_M$, each.

The electric components are obtained by setting $P_M=0$ in
(\ref{bvec}) and solving for $P_E$ while the magnetic components
are obtained by setting $P_E=0$ in (\ref{bvec}) and solving for
$P_M$. Since the resultant wave equation is {\em identical\/} in
both these cases, we shall adopt the generic notation $\Psi$ for
the Debye potentials $P_E$ and $P_M$, both \cite{cohen}.

The physical correspondence with the fields is:\beq E_i &=& f_{i0}
\qquad\qquad B_1 = f_{23} \n \\ B_2 &=& f_{31} \qquad\qquad B_3 =
f_{12} \eeq where the index $i$ ranges from $1, 2, 3$, $E_i$ are
the electric field components and $B_i$ are the magnetic field
components.

Choose (See Appendix - \ref{emdetail}, for computational details)
\be \Psi = T_{\ell n}(t)R_{n}(r)P_{\ell}(\cos{\theta}) e^{im\phi}
\ee where $P_{\ell}(\cos{\theta})$ is an associated Legendre
function, the temporal function $T_{\ell n}(t)$ satisfies
(\ref{twaveeq}) and the radial function $R_n(r)$ satisfies
(\ref{rwaveeq}). As usual, $\ell$ and $m$ are integers: $m$ to
ensure single-valued nature of $e^{im\phi}$ and $\ell$ to ensure
that the associated Legendre functions do not diverge for
$\cos{\theta}=\pm 1$. That is to say, the associated Legendre
functions are polynomials.

When $y'\neq 0$, the radial equation (\ref{rwaveeq}) can be
written as: \be
\frac{d^2R_n}{dy^2}+\frac{1}{y}\frac{dR_n}{dy}+\frac{n}{y^2}R_n =
0\ee and it is an Euler equation. The solutions of this Euler
equation, for $y > 0$, are: \begin{widetext} \beq n>0 \qquad\qquad
R_n &=& c_1
\cos{(\sqrt{n}\ln{y})} + c_2\sin{(\sqrt{n}\ln{y})} \label{ngez} \\
n < 0 \qquad\qquad   R_n &=& c_3 y^{\sqrt{-n}}+c_4
y^{-\,\sqrt{-n}} \label{nlez} \\ n =0 \qquad\qquad   R_n &=& c_5
\ln{y}+c_6 \label{neqz} \eeq where $c_1$, $c_2$, $c_3$, $c_4$,
$c_5$ and $c_6$ are constants.

The temporal equation (\ref{twaveeq}) is of the Fuchsian form and
is amenable to series solutions as per the theorem of Fuchs.
Alternatively, a substitution \be T_{\ell n}(t) = \frac{{\cal
T}_{\ell n}(t)}{\sqrt{B}} \ee can be used to recast
(\ref{twaveeq}) into the form: \be \ddot{{\cal T}}_{\ell n} =
-\,\left( \frac{1}{4}\frac{\dot{B}^2}{B^2}
-\frac{\ell(\ell+1)}{Y^2} -\, \frac{n}{\gamma^2B^2}-
\,\frac{1}{2}\frac{\ddot{B}}{B} \right)\,{\cal T}_{\ell n} \equiv
-\, W(t){\cal T}_{\ell n}=-\,(n_c-n)\frac{{\cal T}_{\ell
n}}{\gamma^2B^2} \label{teqfin} \ee At any given co-moving time,
the temporal function $W(t)$ has an inflexion point $W(t)=0$ at a
critical value denoted by $n_c(\ell, t)$ with \be n_c(\ell, t) =
\frac{\gamma^2B^2}{Y^2}
\left[\frac{Y^2\dot{B}^2}{4B^2}-\frac{Y^2\ddot{B}}{2B}
-\;\ell(\ell+1)\right] \label{nceq} \ee
\end{widetext}
For $n_c >n$, the function $T_{\ell n}(t)$ displays an
oscillatory behavior. On the other hand, for $n_c< n$, the
function $T_{\ell n}(t)$ displays an exponential behavior. At the
critical value, $n=n_c$, the function $T_{\ell n} \propto (\zeta t
+ \iota)/\sqrt{B}$ where $\zeta$ and $\iota$ are constants.

In terms of the solutions $T_{\ell n}(t)$, $R_n(r)$,
$P_{\ell}(\cos{\theta})$ and $e^{im\phi}$, the electric multi-pole
fields, from (\ref{fields}), are:
\begin{widetext}
\beq E_1 &=& \frac{\ell(\ell+1)}{y^2Y^2}T_{\ell
n}(t)R_n(r)P_{\ell}(\cos{\theta}) e^{im\phi}\qquad\qquad B_1 = 0 \n \\
E_2 &=& \frac{1}{\gamma y'ByY}\,T_{\ell n}(t)R_n'(r)\,e^{im\phi}
\frac{d}{d \theta}\,P_{\ell}(\cos{\theta}) \qquad B_2
=\frac{im\csc{\theta}}{y^2Y}\,R_n(r)\,\dot{T}_{\ell
n}(t)\,P_{\ell}(\cos{\theta}) e^{im\phi} \label{emfields}
\\ E_3 &=&\frac{im\csc{\theta}}{\gamma y'ByY} \,T_{\ell
n}(t)R_n'(r)P_{\ell}(\cos{\theta}) e^{im\phi} \qquad\qquad B_3 =
-\,\frac{1}{y^2Y}R_n(r)\dot{T}_{\ell n}(t) e^{im\phi} \frac{d}{d
\theta}\,P_{\ell}(\cos{\theta}) \n \eeq
\end{widetext}

These are the electric multi-poles (except for $\ell =0$), both
static and dynamic. The magnetic multi-poles are obtained
\cite{cohen} similarly from (\ref{waveeq}) and are related to
(\ref{emfields}) by inserting an independent solution $\Psi$ to
(\ref{waveeq}) for $P_M$ and performing the duality operation $E_i
\to B_i$ and $B_i \to -\,E_i$.

For the sake of further works, we explicitly provide here the
magnetic multi-poles. These are:
\begin{widetext}
\beq B_1 &=& \frac{\ell(\ell+1)}{y^2Y^2}T_{\ell
n}(t)R_n(r)P_{\ell}(\cos{\theta}) e^{im\phi}\qquad\qquad E_1 = 0 \n \\
B_2 &=& \frac{1}{\gamma y'ByY}\,T_{\ell n}(t)R_n'(r)\,e^{im\phi}
\frac{d}{d \theta}\,P_{\ell}(\cos{\theta}) \qquad E_2
=-\,\frac{im\csc{\theta}}{y^2Y}\,R_n(r)\,\dot{T}_{\ell
n}(t)\,P_{\ell}(\cos{\theta}) e^{im\phi} \label{mfields}
\\ B_3 &=&\frac{im\csc{\theta}}{\gamma y'ByY} \,T_{\ell
n}(t)R_n'(r)P_{\ell}(\cos{\theta}) e^{im\phi} \qquad\qquad E_3 =
\frac{1}{y^2Y}R_n(r)\dot{T}_{\ell n}(t) e^{im\phi} \frac{d}{d
\theta}\,P_{\ell}(\cos{\theta}) \n \eeq
\end{widetext}
Of particular interest are the ``fall-off properties'' of the
fields. It is noticed that the fields components are proportional
to either $R/y^2$ or $R'/yy'$.

\subsection{Behavior of electromagnetic fields at early times}
At early times, we may consider the initial hyper-surface, at time
$t=t_i$, to contain sparsely distributed, pressure-less matter
collapsing without any radiation. The emission of radiation will
be switched on at some suitable time when the particles of matter
collide and produce some radiation.

Then, as initial condition, we have $B={\rm constant}\equiv B_0$.
Therefore, at $t=t_i$, we have $n_c \approx
-\,\gamma^2B_0^2\,\ell(\ell +1)/Y^2$.

It is then to be noticed that the source-free electromagnetic
fields in this spacetime are to be given as ``initial" conditions.

\subsection{Behavior of electromagnetic fields near the singular hyper-surface}
From (\ref{teqfin}) and (\ref{nceq}), it is clear that $n_c(\ell,
t)$ is an increasing function of $t$ in a collapse situation, {\it
ie}, for $\dot{B}<0$ and $\ddot{B}<0$. Then, any initial mode with
$n_c<n$, {\it ie}, an initially exponential mode, becomes an
oscillatory mode, $n_c>n$, with the progress of gravitational
collapse of matter. The frequency of oscillations of the fields
then continues to increase with the progress of the collapse. In
the limit of the singularity, {\it ie}, for $t\to t_s$ - the
singular hyper-surface, $n_c\to \infty$ and the frequency of field
oscillations becomes infinite.

Thus, all the $\ell$ - modes (electric and magnetic, both), with
$\ell>0$, become oscillatory with the progress of the collapse
irrespective of their nature at initial time. In particular, this
is the case near the singular hyper-surface $t=t_s$. Note that
this is happening only asymptotically with time. The singular
hyper-surface is to the infinite future of a co-moving observer
who is also the ``cosmological observer" for the spacetime of
(\ref{ssmetfinal}).

This is clearly consistent with the result that the co-moving
observer too becomes relativistic in the limit $t \to t_s$.
Therefore, the co-moving observer only sees radiation in this
limit. That the co-moving observer becomes relativistic in the
limit of the singular hyper-surface is irrespective of whether
there is any black hole - the infinite red-shift surface - in the
spacetime of (\ref{ssmetfinal}). This is also irrespective of
whether there are electromagnetic fields in the spacetime of
(\ref{ssmetfinal}) or not.

However, this is not the astrophysically interesting or relevant
situation since it is reached for an infinite co-moving time when
matter in the entire spacetime has attained relativistic speeds as
a result of the continued pile up of matter to the center of the
spacetime.

\subsection{Behavior of electromagnetic fields in the presence of a black hole}
Perhaps, what is astrophysically relevant is the situation of a
black hole existing in the spacetime of (\ref{ssmetfinal}).
However, it must be emphasized that a black hole is only an
infinite red-shift surface here and that it must exist in the
spacetime of (\ref{ssmetfinal}) as an ``initial" condition for
non-singular and non-degenerate matter data. It is therefore
likely that the black hole of (\ref{ssmetfinal}) may only be of
academic interest. That is to say, the black hole of
(\ref{ssmetfinal}) may not be astrophysically relevant.

Then, of some, nonetheless questionable, astrophysical interest is
the behavior of electromagnetic fields when a black hole exists in
(\ref{ssmetfinal}) as an infinite red-shift surface.

As seen in \S\ref{bhole}, the necessary condition for the presence
of a black hole in (\ref{ssmetfinal}) is $y(r) < 1$. Therefore, we
need to consider the field solutions (\ref{emfields}) in the range
$y:(y_c, \infty)$ when $y_c < 1$. We will then obtain the fields
within the black hole region for $y:(y_c, 1)$ and outside or
exterior to the black hole region for $y:(1, \infty)$.

In this case, the radial behavior is obtained for values $y_c<y<1$
and there is no difference in the temporal behavior of the fields.

\section{Discussion} \label{discussion}
In General Relativity, a {\em continuum\/} of ``curved"
4-dimensional spacetime geometry describes the evolution of
matter. In its (3+1)-formulation, we can consider some
distribution of gravitating matter on an ``initial" spacelike
hyper-surface, the Cauchy surface, and, from the Einstein
equations, can obtain the temporal evolution of matter from that
``initial" datum.

We may select a variety of ``matter datum" on the initial
hyper-surface. For example, we may consider a single particle,
{\it ie}, a single mass-point, or a blob of matter surrounded by
vacuum or a blob of matter that is radiating, and evolve that
matter datum using the field equations.

Now, any such ``initial" data is {\em replaceable\/} with ``that"
mass-point or ``that" blob of matter being surrounded by ``more"
matter, this replacement being {\em ad infinitum\/} till the
entire initial hyper-surface has matter everywhere. Note also that
this ``replacement" can be achieved in uncountably many different
ways, {\it ie}, by distributing ``more" matter in uncountably many
ways on the initial hyper-surface.

Further, each stage of the evolution is obtainable from the
previous stage of matter. This is the principle of causal
development.

Now, when the matter datum is specified over all of the initial
hyper-surface, we obtain a ``cosmological" situation or spacetime.

Different initial data could evolve to {\em distinct\/}
four-dimensional spacetime geometries. Therefore, these spacetime
geometries; of a mass-point, of a matter blob and the
``cosmological" spacetime obtained for an {\em ad infinitum\/}
replacement of any of the considered situations; possess different
geometric and, hence, physical features.

The question then arises: Which geometric or physical features of
these spacetimes are {\em relevant\/} to ``real" objects of the
observed Universe? To be able to make contact with physical
objects embedded in the Universe, a spacetime of the object in
question is then needed to be ``cosmological".

Thus, a point of view can be advocated that the geometric features
of only cosmological spacetimes are the ones that are relevant to
actual physical objects of the real Universe.

The point here is that, in General Relativity, the idealization of
an ``isolated" object comes with its own pitfalls of the above
nature. The issue here is that we could, without changing the
Newtonian law of gravitation, add two masses to produce a new mass
in the Newtonian theory but that is not permissible with
``spacetimes" of arbitrary nature in General Relativity.

Note that the local spacetime geometry is, of course, Minkowskian
since the cosmological spacetime is locally flat. Moreover, it may
also be that some features of a cosmological spacetime are present
with non-cosmological spacetimes.

In the above general spirit, we discussed, in this paper,
``physical characteristics" of a radially homothetic spacetime of
(\ref{ssmetfinal}).

We first showed that the requirement of radial homothety, {\it
ie}, of the existence of a radial homothetic Killing vector for a
spherical spacetime, fixes the spacetime metric uniquely to
(\ref{ssmetfinal}). The existence of a radial HKV allows, in
accordance with Lie's theory, an arbitrary function of the
co-moving radial coordinate in the metric.

In a radially homothetic spacetime of (\ref{ssmetfinal}), there
is, therefore, the maximal ``physical" freedom, but not the
geometrical freedom, compatible with the assumption of spherical
symmetry. Thus, we concentrated on the non-singular and
non-degenerate matter data for (\ref{ssmetfinal}).

We then discussed the issue of the regularity of the center of a
radially homothetic spacetime and discussed the behavior of
physical quantities. The lack of regularity of the center of
(\ref{ssmetfinal}) for non-singular and non-degenerate data is
compatible with the ``physical" expectation that the
``cosmological" observer does not witness the formation of the
black hole.

Next, we showed that a black hole  of (\ref{ssmetfinal}) is only
an infinite red-shift surface and is a part of the ``initial" data
of the spacetime.

Further, the curvature singularity of such a spacetime is a
singular spacelike hyper-surface at which matter in the spacetime
attains the speed of light. This spacetime singularity of
(\ref{ssmetfinal}) occurs to the future of every observer in
(\ref{ssmetfinal}). Hence, the spacetime singularity of
(\ref{ssmetfinal}) is not visible to any observer. That is to say,
the singularity of (\ref{ssmetfinal}) is not naked locally or
globally.

These results are in agreement with the strong Cosmic Censorship
that demands that the spacetime singularity be not visible to any
observer unless and until it is actually encountered
\cite{penrose98}.

We then discussed the steps of collapse of matter from the initial
dusty state. It was argued that ``all" the necessary ingredients
of the expected physical evolution of matter are obtainable in the
radially homothetic spacetime.

Following the Cohen-Kegeles generalization \cite{cohen} of the
flat-space Hertz-Debye potential formalism, we have also obtained
the source-free electromagnetic fields, (\ref{emfields}), in the
spacetime of (\ref{ssmetfinal}). In the limit of the singular
hyper-surface, all the field modes are oscillatory. This behavior
of the electromagnetic fields is asymptotically reached in the
limit of infinite co-moving time.

In this spacetime, when self-gravity dominates with finality, the
unstoppable collapse begins and leads to an eventual spacetime
singularity, the singular hyper-surface,  only asymptotically.
However, such a situation is not astrophysically relevant since
``all" the matter in the spacetime becomes relativistic in this
situation. This is expected only in the asymptotic future.

The radially homothetic spacetime of (\ref{ssmetfinal}) therefore
provides us the spacetime of {\em astrophysically interesting}
gravitational collapse problem. In many such collapse situations,
matter may trap and carry radiation with it as it collapses.

From the astrophysical point of view, the recent observations
appear to point to the existence of a ``null hyper-surface" in
candidate objects. In particular, \cite{narayan1, narayan2} have
recently pointed out that potential black hole candidates and
known neutron stars separate in two categories in
$\log{(L_{max})}$ versus $\log{(L_{min}/L_{max})}$ plots for X-ray
luminosity of these sources.

The explanation for this separation is provided on the basis of
the Advection-Dominated Accretion Flows (ADAF) that have more core
luminosity in the X-rays in the case of Neutron Stars. The
accreting matter exhibiting energy-advection encounters the
physical surface of the neutron star at which it deposits the
stored energy to become X-ray bright. On the other hand, for a
black hole, matter is expected to encounter no such physical
surface and, hence, matter is not expected to become X-ray bright.

The point is that in these models of ADAF the above is implemented
in the form of a boundary condition that essentially implies that
the matter exhibiting energy-advection accretes without
encountering a physical surface like that of a neutron star or
not.

The black hole of (\ref{ssmetfinal}) - an infinite red-shift
surface - is, in the co-moving frame, approached asymptotically by
collapsing matter that crosses it in its rest frame. Consequently,
matter falling onto the black hole will appear, to a co-moving
observer, to be collapsing without encountering a physical surface
like that of a neutron star. Hence, matter displaying
energy-advection will not deposit the stored energy at any
surface. Such an object will, therefore, appear less X-ray bright
as compared to an object with a physical surface. This is as per
the interpretation of observations in \cite{narayan1, narayan2}.

However, the black hole of (\ref{ssmetfinal}) is a part of the
``initial" data for it. Therefore, unless a black hole is assumed
to exist, there is no reason for the non-existence of any hard
surface of the collapsing object in the spacetime of
(\ref{ssmetfinal}).

Thus, it seems that the reason(s) for the observations used in
\cite{narayan1, narayan2} are, in all probability, different than
have been explored therein. For (\ref{ssmetfinal}), such reasons
can only be explored on the basis of the detailed ``physical"
considerations leading to solutions to (\ref{density}) and
(\ref{pressure}).

In (\ref{ssmetfinal}), any spherical ``object" will collapse
asymptotically to the singular hyper-surface. It is an Eternally
Collapsing Object. However, collapse to singular hyper-surface of
(\ref{ssmetfinal}) requires the entire spacetime to be
``relativistic" and that takes infinite co-moving time. Therefore,
in (\ref{ssmetfinal}), the ``Eternally Collapsing Object" is also
not relevant to the present astrophysical observations that have
been used in \cite{narayan1, narayan2}.

Then, from the results of the present work, it seems most likely
that the astrophysically relevant possibility is that of
``sufficiently Collapsed Objects" accreting matter in
astrophysical environments. We are, of course, considering only
spherically symmetric situation in this paper.

Now, from the above, it is then puzzling that no null membrane
black holes emerged in the present study. The following issue,
therefore, arises.

\subsection*{Issue of null membranes in the {\em physically realizable\/}
collapse}

The standard scenario of gravitational collapse leading to a null
membrane black hole is, so far, only an ``expectation" based on
plausibility arguments. The exact general relativistic spacetime
of the gravitational collapse of a ``physical" star supporting the
standard scenario with a null membrane is not known.

We have shown here that the radially homothetic, spherical
spacetime of (\ref{ssmetfinal}) describes the {\em physically
realizable\/} spherical collapse.

Therefore, the standard picture of spherical collapse of matter
leading to a null membrane black hole can be verified using
(\ref{ssmetfinal}).

We have then shown that a null membrane does not arise for {\em
physically realizable\/} gravitational collapse from any
non-singular and non-degenerate data in (\ref{ssmetfinal}).
However, as has been shown here, an infinite red-shift surface can
``exist" in (\ref{ssmetfinal}) as a part of non-singular and
non-degenerate data.

The standard picture of the gravitational collapse leading to a
null membrane black hole cannot therefore be realized in this
spacetime for non-singular and non-degenerate data.

Then, the question is of the existence of {\em another}
cosmological spacetime describing {\em physically realizable\/}
spherical collapse of matter and supporting the standard
expectation of the existence of a null membrane black hole.

The question will also be of the difference between such a
spacetime and a radially homothetic spacetime considered here.
Does general relativity allow {\em two\/} inequivalent such
spherical spacetimes? This is a fundamental issue.

\acknowledgements I am grateful to Ravi Saraykar, Keshlan Govinder
and Pradeep Muktibodh for verifying the calculations and for
helpful discussions. Some of the reported calculations have been
performed using the software {\tt SHEEP} and I am indebted to
Malcolm MacCallum for providing this useful package and for a
helpful comment about the type-D nature of the spacetime
considered here.

\appendix
\begin{widetext} \section{Computational Details}
\label{compudetail}
\subsection{Electromagnetic fields} \label{emdetail}
Some useful Hodge dual operations on basis forms are:
                                            \bigskip

\noindent
\begin{tabular}{llll} $\star\,
\omega^0\wedge\omega^1\wedge\omega^2\wedge\omega^3 = {\bf
1}$\hspace{.7in} &$ \star \,{\bf 1} = -
\,\omega^0\wedge\omega^1\wedge\omega^2\wedge\omega^3$\hspace{.7in}
& $\star \,\omega^0 \wedge \omega^1 = \omega^2\wedge\omega^3$  \\ \\
$\star \,\omega^0 \wedge \omega^2 = -\, \omega^1\wedge\omega^3$ &
$\star \,\omega^0 \wedge \omega^3 = \omega^1\wedge\omega^2$ &
$\star \,\omega^1 \wedge \omega^2 = -\, \omega^0\wedge\omega^3$
\\ \\ $\star \,\omega^1 \wedge \omega^3 = \, \omega^0\wedge\omega^2$
& $\star \,\omega^2 \wedge \omega^3 = -\,\omega^0\wedge\omega^1$ &
$\star \,\omega^1\wedge\omega^2\wedge\omega^3 = \omega^0$ \\ \\
$\star \,\omega^0\wedge\omega^2\wedge\omega^3 = \omega^1$ & $\star
\,\omega^0\wedge\omega^3\wedge\omega^1 = \omega^2$ & $\star
\,\omega^0\wedge\omega^1\wedge\omega^2 = \omega^3$ \\ \\ $\star
\,\omega^0 = \omega^1\wedge\omega^2\wedge\omega^3$ & $\star
\,\omega^1 = \omega^0\wedge\omega^2\wedge\omega^3$ & $\star
\,\omega^2 = \omega^0\wedge\omega^3\wedge\omega^1$ \\ \\ & $\star
\,\omega^3 = \omega^0\wedge\omega^1\wedge\omega^2$
\end{tabular} \bigskip

\noindent Now, in the following we shall set $P_M=0$ in
(\ref{bvec}), that is to say, we shall evaluate the electric
multi-poles for the metric (\ref{ssmetfinal}). We shall also
denote the potential as $\Psi$. Then, from (\ref{bvec}), we obtain
\beq P &=& \Psi\, \omega^0 \wedge \omega^1 \\
\star\,P &=& \Psi\, \omega^2\wedge\omega^3 =
\Psi\,y^2Y^2\sin{\theta}\,d\theta\wedge d\phi \n \\
d\,\star\,P &=& \left(
\Psi_{,t}+2\Psi\frac{\dot{Y}}{Y}\right)y^2Y^2\sin{\theta}\,
dt\wedge d\theta \wedge d\phi +\left(
\Psi_{,r}+2\Psi\frac{y'}{y}\right)y^2Y^2\sin{\theta}\,dr\wedge
d\theta \wedge d\phi \n
\\ &=& \left( \frac{\Psi_{,t}}{y}+\frac{2\Psi}{y}\frac{\dot{Y}}{Y}\right)
\omega^0\wedge\omega^2\wedge\omega^3 + \left(
\frac{\Psi_{,r}}{\gamma y'B}+\frac{2\Psi}{\gamma yB}\right)
\omega^1\wedge\omega^2\wedge\omega^3 \n \\ \star\,d\,\star\,P &=&
\left( \frac{\Psi_{,t}}{y}+\frac{2\Psi}{y}\frac{\dot{Y}}{Y}\right)
\omega^1 + \left( \frac{\Psi_{,r}}{\gamma y'B}+\frac{2\Psi}{\gamma
yB}\right) \omega^0 \n \\ &=& \left( \frac{\gamma
y'B\Psi_{,t}}{y}+\frac{2\Psi \gamma y'B \dot{Y}}{yY}\right) \,dr +
\left( \frac{y\Psi_{,r}}{\gamma y'B}+\frac{2\Psi}{\gamma B}\right)
\,dt \n \\ d \star d\star P &=& \left[ \left(
B\Psi_{,tt}+\dot{B}\Psi_{,t} +2\Psi_{,t}B\frac{\dot{Y}}{Y} + 2\Psi
B\frac{\ddot{Y}}{Y} -2\Psi B \frac{\dot{Y}^2}{Y^2} + 2\Psi
\dot{B}\frac{\dot{Y}}{Y}\right)\frac{\gamma y'}{y}\right. \n \\
&\phantom{=}&\left.  -\left( \frac{y\Psi_{,rr}}{y'} +
3\Psi_{,r}-\frac{yy''\Psi_{,r}}{y'^2} \right) \frac{1}{\gamma
B}\right] dt\wedge dr -\left( \frac{y\Psi_{,r\theta}}{\gamma y' B}
+
\frac{2\Psi_{,\theta}}{\gamma B}\right) dt \wedge d\theta \n \\
&\phantom{=}&-\left( \frac{y\Psi_{,r\phi}}{\gamma y' B} +
\frac{2\Psi_{,\phi}}{\gamma B}\right) dt \wedge d\phi  -\left(
\Psi_{,t\theta} +
2\Psi_{,\theta}\frac{\dot{Y}}{Y}\right)\frac{\gamma y' B}{y} dr \wedge d\theta \n \\
&\phantom{=}& -\left( \Psi_{,t\phi} +
2\Psi_{,\phi}\frac{\dot{Y}}{Y}\right)\frac{\gamma y' B}{y} dr
\wedge d\phi \n \eeq \beq d \star d\star P &=& \left[ \left(
B\Psi_{,tt}+\dot{B}\Psi_{,t} +2\Psi_{,t}B\frac{\dot{Y}}{Y} + 2\Psi
B\frac{\ddot{Y}}{Y} -2\Psi B \frac{\dot{Y}^2}{Y^2} + 2\Psi
\dot{B}\frac{\dot{Y}}{Y}\right)\frac{\gamma y'}{y}\right. \n \\
&\phantom{=}&\left.  -\left( \frac{y\Psi_{,rr}}{y'} +
3\Psi_{,r}-\frac{yy''\Psi_{,r}}{y'^2} \right) \frac{1}{\gamma
B}\right]\frac{1}{\gamma y'By} \omega^0\wedge \omega^1 -\left(
\frac{\Psi_{,r\theta}}{y'} +
\frac{2\Psi_{,\theta}}{y}\right)\frac{1}{\gamma ByY} \omega^0 \wedge \omega^2 \n \\
&\phantom{=}&-\left( \frac{\Psi_{,r\phi}}{y'} +
\frac{2\Psi_{,\phi}}{y}\right)\frac{\csc{\theta}}{\gamma ByY}
\omega^0 \wedge \omega^3 -\left( \Psi_{,t\theta} +
2\Psi_{,\theta}\frac{\dot{Y}}{Y}\right)\frac{1}{y^2Y} \omega^1 \wedge \omega^2 \n \\
&\phantom{=}& -\left( \Psi_{,t\phi} +
2\Psi_{,\phi}\frac{\dot{Y}}{Y}\right)\frac{\csc{\theta}}{y^2Y}
\omega^1 \wedge \omega^3 \n \eeq Similarly, \beq P&=& \Psi
\omega^0\wedge \omega^1 = \Psi yy'\gamma B dt\wedge dr \\ dP &=&
\Psi_{,\theta}\, yy'\gamma B\, d\theta\wedge dt\wedge dr +
\Psi_{,\phi}\, yy'\gamma B\, d\phi\wedge dt\wedge dr =
\frac{\Psi_{,\theta}}{yY}\,\omega^2\wedge\omega^0\wedge\omega^1 +
\frac{\csc{\theta}\Psi_{,\phi}}{yY}\, \omega^3\wedge\omega^0
\wedge\omega^1 \n \\ \star\,dP &=&
\frac{\Psi_{,\theta}}{yY}\,\omega^3-\frac{\csc{\theta}\Psi_{,\phi}}{yY}\,\omega^2
= \Psi_{,\theta}\,\sin{\theta}\,d\phi- \csc{\theta}\Psi_{,\phi}
\,d\theta \n \\ d\star\,dP &=& \sin{\theta}\left[ \Psi_{,\theta
t}\, dt \wedge d\phi + \Psi_{\theta r}\, dr\wedge d\phi +
\left( \Psi_{,\theta\theta}+\Psi_{,\theta}\,\cot{\theta}\right) \, d\theta\wedge d\phi \right] \n \\
&\phantom{=}& \qquad\qquad\qquad - \csc{\theta}\left( \Psi_{,\phi
t}\, dt \wedge d\theta + \Psi_{,\phi r}\, dr \wedge d\theta
+\Psi_{,\phi \phi}\, d\phi \wedge d\theta \right)\n \\ &=&
\frac{\Psi_{,\theta t}}{y^2Y}\,\omega^0\wedge\omega^3 +
\frac{\Psi_{,\theta r}}{\gamma y' ByY}\,\omega^1\wedge\omega^3 +
\left( \Psi_{,\theta\theta}+\Psi_{,\theta}\cot{\theta}+
\csc^2{\theta} \Psi_{,\phi\phi}\right) \frac{1}{y^2Y^2}
\,\omega^2\wedge\omega^3 \n \\
&\phantom{=}& \qquad\qquad\qquad -\frac{\csc{\theta}\Psi_{,\phi
t}}{y^2Y} \, \omega^0\wedge\omega^2 -
\frac{\csc{\theta}\Psi_{,\phi r}}{\gamma y'ByY} \,
\omega^1\wedge\omega^2 \n \\ \star\,d\,\star\,dP &=& -\left[
\Psi_{,\theta\theta}+\cot{\theta}\,\Psi_{,\theta}+\csc{\theta}\,\Psi_{,\phi\phi}
\right]\frac{1}{y^2Y^2}\,\omega^0\wedge\omega^1 +
\frac{\Psi_{,\theta r}}{\gamma y'ByY}\, \omega^0\wedge\omega^2  +
\frac{\csc{\theta}\Psi_{,\phi r}}{\gamma y'ByY}\,
\omega^0\wedge\omega^3 \n \\ &\phantom{=}& \qquad\qquad +
\frac{\csc{\theta}\Psi_{,\phi t}}{y^2Y}\, \omega^1\wedge\omega^3 +
\frac{\Psi_{,\theta t}}{y^2Y}\, \omega^1\wedge\omega^2 \n \eeq
Therefore, \beq \Delta P &=& \left\{\phantom{\frac{\dot{Y}}{Y}}
\left(B\Psi_{,tt}+\dot{B}\Psi_{,t} +2\Psi_{,t}B\frac{\dot{Y}}{Y} +
2\Psi B\frac{\ddot{Y}}{Y} -2\Psi B \frac{\dot{Y}^2}{Y^2} + 2\Psi
\dot{B}\frac{\dot{Y}}{Y}\right)\frac{1}{y^2B} \right. \n \\
&\phantom{=}&  \left. -\left( \frac{y\Psi_{,rr}}{y'} +
3\Psi_{,r}-\frac{yy''\Psi_{,r}}{y'^2} \right) \frac{1}{\gamma^2
B^2y'y} -\left(
\Psi_{,\theta\theta}+\cot{\theta}\,\Psi_{,\theta}+\csc{\theta}\,\Psi_{,\phi\phi}
\right)\frac{1}{y^2Y^2}\phantom{\frac{\dot{Y}}{Y}}\right\}
\omega^0\wedge\omega^1 \n \\ &\phantom{=}& -
\frac{2\Psi_{,\theta}}{\gamma By^2Y}\,\omega^0\wedge\omega^2 -
\frac{2\csc{\theta}\Psi_{,\phi}}{\gamma
By^2Y}\,\omega^0\wedge\omega^3 -
\frac{2\dot{Y}\Psi_{,\theta}}{y^2Y^2}\,\omega^1\wedge\omega^2 -
\frac{2\csc{\theta}\dot{Y}\Psi_{,\phi}}{y^2Y^2}\,\omega^1\wedge\omega^3
\label{finwave} \eeq
\bigskip

\noindent Now, with the gauge term as \be G= \frac{2\Psi}{\gamma
By}\,\omega^0 + \frac{2\dot{Y}\Psi}{yY}\,\omega^1\ee we get \beq
dG &=& \left[\left( \frac{2\ddot{Y}\Psi B}{yY}
+\frac{2\dot{Y}\dot{\Psi}B}{yY} - \frac{2\Psi\dot{Y}^2B}{yY^2} +
\frac{2\Psi\dot{Y}\dot{B}}{yY}\right)\frac{1}{By} -
\frac{2\Psi_{,r}}{\gamma^2B^2y'y} \right]\,\omega^0\wedge\omega^1
\n \\ &\phantom{=}& - \frac{2\Psi_{,\theta}}{\gamma
By^2Y}\,\omega^0\wedge\omega^2 -
\frac{2\csc{\theta}\Psi_{,\phi}}{\gamma
By^2Y}\,\omega^0\wedge\omega^3 -
\frac{2\dot{Y}\Psi_{,\theta}}{y^2Y^2}\,\omega^1\wedge\omega^2 -
\frac{2\csc{\theta}\dot{Y}\Psi_{,\phi}}{y^2Y^2}\,\omega^1\wedge\omega^3
 \label{fingauge} \eeq

Therefore, substituting (\ref{finwave}) and (\ref{fingauge}) in
(\ref{waveeq}), it is seen that the $\omega^0\wedge\omega^1$ term
yields the required wave equation while all other terms lead to
identities. The wave equation is: \beq Y^2\left(
\Psi_{,tt}+\frac{\dot{B}}{B}\Psi_{,t}\right) - \left(
\frac{y\Psi_{,rr}}{y'}+\Psi_{,r}-\frac{yy''\Psi_{,r}}{y'^2}\right)
\frac{y}{y'}\,\frac{Y^2}{\gamma^2B^2} =
\Psi_{,\theta\theta}+\cot{\theta}\,\Psi_{,\theta} +\csc{\theta}\,
\Psi_{,\phi\phi} \label{fwaveeq} \eeq and is amenable to solution
by separation of variables. Dividing (\ref{fwaveeq}) by $\Psi$ and
setting \be \Psi\left( t, r, \theta, \phi \right) = T_{\ell n}
(t)\,R_n(r)\,P_{\ell}(\cos{\theta})\, e^{im\phi} \ee where
$P_{\ell}(\cos{\theta})$ is an associated Legendre polynomial, we
obtain \beq \frac{\ddot{T}_{\ell n}}{T_{\ell n}
}+\frac{\dot{B}}{B}\frac{\dot{T}_{\ell n}}{T_{\ell n}} +
\frac{\ell(\ell +1)}{Y^2}+\frac{n}{\gamma^2B^2} &=& 0 \label{twaveeq} \\
\frac{R_n''}{R_n} + \left( \frac{y'}{y} -
\frac{y''}{y'}\right)\,\frac{R_n'}{R_n}+
n\,\left(\frac{y'}{y}\right)^2 &=& 0 \label{rwaveeq} \eeq where
$m$, $\ell$ and $n$ are separation constants. In the above, an
overhead dot denotes a time-derivative and an overhead prime
denotes a derivative with respect to $r$.

\end{widetext}

\end{document}